\documentstyle[12pt]{article}
\newcommand{\mbf}[1]{\mbox{\boldmath $ #1 $\unboldmath}}
\def\zhat{{\rm\bf e}_z}
\def\be{\begin{equation}}
\def\ee{\end{equation}}
\def\bea{\begin{eqnarray}}
\def\eea{\end{eqnarray}}
\def\ddr{\frac{\partial}{\partial r}}
\def\ddz{\frac{\partial}{\partial z}}
\def\ddt{\frac{\partial}{\partial t}}
\def\ddzs{\frac{\partial}{\partial z_*}}
\def\ddts{\frac{\partial}{\partial t_*}}

\def\eps{\Delta\kappa}
\def\bp{{\beta_+}}
\def\bm{{\beta_-}}
\def\ky{{\kappa_Y}}
\def\ke{{\kappa_{\rm eq}}}
\begin{document}

\begin{titlepage}
\vspace*{-62pt}
\begin{flushright}
FERMILAB-Pub-95/074-A\\
astro-ph/9504051\\
April 1995\\
Appearing in {\it Journal of Fluid Mechanics} {\bf 312},327-340.
\end{flushright}
\vspace{1.5cm}
\begin{center}
{\Large \bf Ekman pumping in compact astrophysical bodies}\\
\vspace{.6cm}
\small
{\bf By MARK ABNEY$^{1,2}$ \& RICHARD I.  EPSTEIN$^2$}
\vspace{12pt}

{

$^1$Department of Astronomy and Astrophysics, The Enrico Fermi
Institute,\\
The University of Chicago, Chicago, Illinois~~~60637\\
and\\
NASA/Fermilab Astrophysics Center,\\
Fermi National Accelerator Laboratory, Batavia, Illinois~~~60510
\vspace{12pt}

$^2$Los Alamos National Laboratory, \\
MS D436, Los Alamos, New Mexico~~~87545
\vspace{18pt}
}
\normalsize
\end{center}

\vspace*{12pt}

\begin{quote}
\hspace*{2em}
We examine the dynamics of a rotating viscous fluid following an
abrupt change in the angular velocity of the solid bounding surface. We
include the effects of a
density stratification and compressibility which are important in astrophysical
objects such as neutron stars.
We confirm and extend the conclusions of previous studies that
stratification restricts the Ekman pumping process to a
relatively thin layer near the boundary, leaving much
of the interior fluid unaffected. We find that finite compressibility further
inhibits Ekman pumping by decreasing the extent of the pumped layer and
by increasing the time for spin-up. The results of this
paper are important for interpreting the spin period discontinuities
(``glitches'') observed in rotating neutron stars.

\vspace*{12pt}


\vspace*{12pt}

\noindent
\small email: $^{1,2}$abney@oddjob.uchicago.edu
\hspace*{1.5cm}
$^2$epstein@lanl.gov

\end{quote}


\end{titlepage}

\baselineskip=24pt

\section{Introduction}

The approach to solid-body rotation of a fluid inside a rotating
boundary is a familiar phenomenon with many 
applications. For instance, not only can we directly observe
this phenomenon in the laboratory, but it may also play an
important role in solar models, neutron stars and other
environments.
Greenspan \& Howard (1963) give a fundamental
analysis of the linearized version of this problem by
considering a rotating axisymmetric container filled
with a viscous incompressible fluid. They
examine the behaviour of the fluid after the
angular velocity of the container is suddenly changed by a small
amount.  Their solution consists of
three  distinct, time-separated phases: boundary layer
formation, Ekman pumping and viscous relaxation. 
The bulk of
the fluid spin-up (or down) occurs
through Ekman pumping. Subsequent studies
of the effect a stratification in density has
on the Ekman pumping process (Walin 1969; Sakurai 1969; Buzyna \&
Veronis 1971; Hyun 1983; Spence, Foster \& Davies 1992)
found that some regions of the interior do not reach the final
angular velocity until the viscous diffusion time has
elapsed. The basic three stage structure of spin-up, however,
remains intact where the intermediate Ekman pumping time results
in a quasi-steady state in the fluid interior (see the review of
Benton \& Clark 1974 for a complete discussion of this subject).

The spin-up of a compressible fluid has been studied in the
context of rapidly rotating gases (Sakurai \& Matsuda 1974;
Bark, Meijer \& Cohen 1978; Lindblad, Bark, \& Zahrai 1994). These
studies concluded that the basic structure for spin-up remained
unchanged, but with some important and interesting differences.
Numerical studies have also been
carried out for both the linear and non-linear regimes (Hyun \&
Park 1990, 1992; Park \& Hyun 1994) with the results lending support to the
analytical results. The compressible fluid
research, however, was motivated by a desire to understand the
dynamics of gas centrifuges where the effects of gravity are
negligible compared to centrifugal forces. A density gradient
results, but in a direction perpendicular to the angular
velocity. 

In this investigation, we are primarily concerned with the
possible astrophysical 
applications of the theory of stratified, compressible rotating
fluids. Indeed one of the primary motivations for studying
spin-up of a stratified fluid was to understand the sun's
rotation (Howard, Moore \& Spiegel 1967; Sakurai, Clark \& Clark
1971; Clark, Clark, Thomas \& Lee 1971). Though the sun does not
have a solid outer crust, Ekman suction may arise because of
solar wind torque, or boundary layer flow
taking place between fluid interfaces. Ekman pumping in
multilayer fluids was investigated by Pedlosky (1967) with his
theory tested experimentally by Linden \& van Heijst (1984) and
O'Donnell \& Linden (1992).
Recently, the possibility that Ekman pumping may play a role in
the synchronization of some binary stars has also been discussed
(Tassoul \& Tassoul 1990, 1992; Rieutord 1992). We, however, are
motivated by the phenomenon of pulsar ``glitches,'' a sudden
slight increase in a neutron star's rotational frequency, and the
resultant response of the fluid core. The
simplified model of a rotating neutron star we consider includes
the effect of compressibility of the inner fluid as well as a
density gradient parallel to the gravitational field and angular
velocity. A strong 
magnetic field, as exists on the surface, may alter the internal
dynamics, but the presence of a magnetic field in the core is
not well established and we choose to ignore it. The effects of a
magnetic field which  threads the core are discussed in greater
detail in a forthcoming paper.


\section{Ekman pumping}

\begin{center}
  {\it 2.1 Fluid dynamics}
\end{center}

To investigate the response of the fluid in a
rotating container, we examine the usual, simple model of a
cylinder of height $2L_*$ and radius $r_{c*}$
rotating with angular velocity ${\bf\Omega}_*$  (here and elsewhere
an asterisk subscript indicates a dimensional variable or operator; 
quantities without this subscript are dimensionless.) 
When the angular velocity of the container is abruptly
changed by a small amount, the differential rotation between the
fluid and the top and bottom of the cylinder generates the
``Ekman pumping'' process.
Unlike  previous studies, we introduce an equation of state
relating the mass-energy density $\rho_*$ to the pressure
$p_*$ and to the composition.
For a  fluid with a kinematic viscosity $\nu_*$, the
Navier--Stokes equations 
of motion in a frame rotating with angular velocity ${\bf
\Omega}_*$ are  
\be
    \rho_*\left (\frac{\partial{\bf v}_*}{\partial t_*}+{\bf v}_*
   \cdot\mbf{ \nabla}_* {\bf v}_*
 +2({\bf \Omega}_*\times {\bf v}_*)\right )
  =-\mbf {\nabla}_* p_*+\rho_*{\bf g}_*+\frac{1}{2}
  \rho_*\mbf{ \nabla}_*\Omega_*^2r_*^2
  +\rho_*\nu_*\nabla_*^2{\bf v}_*, 
\ee
where ${\bf g}_*$ is,
\be
  {\bf g}_*= \left\{ \begin{array}{ll}
                                   -g_*\zhat & z_*>0 \\
				   +g_*\zhat & z_*<0
		\end{array} \right. ,	  
\ee
$r_*$ is the cylindrical
radius, and we take ${\bf \Omega}_*=\Omega_*\zhat$ with $\zhat$
the unit vector in the $z$-direction.
{}From this point onward we
consider only the upper half--plane, noting that all quantities
are symmetric about $z_*=0$.

This particular form of ${\bf g}_*$ represents a constant
inward pointing acceleration, even though in a self--gravitating
body the 
acceleration is a decreasing function of the radius, becoming
zero at the center. However, as we will show below,
for the parameter ranges applicable to a neutron
star the significant dynamics occurs  in a thin layer near the
boundaries where
gradients to the gravitational acceleration are  negligible.

As long as $r_{c*}$ is not too large the centrifugal acceleration
is small compared to the gravitational acceleration and can be
neglected.  More precisely, we assume that
finite Froude number effects can be ignored, i.e.\ 
$F\equiv4\Omega_*^2r_{c*}/g_*\ll1$, where $F$ is the Froude number.
This results in a state of rotational equilibrium
where the pressure $p_{s*}$ and density
$\rho_{s*}$ are functions only of $z_*$.
The Navier--Stokes equation for the equilibrium system is
\be
  \ddzs p_{s*}=-\rho_{s*} g_*. \label{equil}
\ee

We now look at a perturbed system in which the
angular velocity of the boundary is suddenly changed by a
relatively small 
amount $\Delta\Omega_*$. The resulting  pressure and
density are
\bea
  p_*&=&p_{s*}(z_*)+\delta p_*(r_*,z_*,t_*) \\
  \rho_* &=& \rho_{s*}(z_*) + \delta\rho_*(r_*,z_*,t_*) .
\eea
To first order in
$\delta p_*$, $\delta\rho_*$ and $v_*(r_*,z_*,t_*)$ we have
\be
  \frac{\partial{\bf v}_*}{\partial t_*} +2\Omega_*\zhat\times {\bf v}_*
   =-\frac{1}{\rho_{s*}}\mbf {\nabla}_*\delta p_*-\frac{1}{\rho_{s*}}
    \delta\rho_*\,{\bf g}_*
    +\nu_*\nabla_*^2{\bf v}_*.
\ee
We non-dimensionalize the equations
by writing variables and operators as a dimensional constant times a
non-dimensional variable or operator as follows,
\begin{eqnarray*}
  {\bf v}_*&\equiv&(L_*\Delta\Omega_*)\,{\bf v}\\
  t_*&\equiv&(E^{1/2}2\Omega_*)^{-1}\, t\\
  {\bf r}_*&\equiv&L_*\,{\bf r}\\
  {\bf z}_*&\equiv&L_*\,z\,\zhat\\
  \delta p_*&\equiv&(2\Omega_*\rho_{0*} L_*^2\Delta\Omega_*) \delta p\\
  \delta\rho_* &\equiv&(2\Omega_*\rho_{0*}L_*\Delta\Omega_*/g_*)\delta\rho\\
  \rho_{s*}&\equiv&\rho_{0*}\,\rho_{s}\\
  \mbf{\nabla}_*&\equiv&(1/L_*)\mbf{\nabla}
\end{eqnarray*}
where $\rho_{0*}$ is a fiducial value for the equilibrium
density. 
We also introduce the dimensionless viscosity, or Ekman number,
\be
  E=\frac{\nu_*}{2\Omega_*L_*^2}.
\ee
The Navier--Stokes equation for the perturbations is now
\be
  E^{1/2}\frac{\partial{\bf v}}{\partial t} +\zhat\times {\bf v}
   =-\frac{1}{\rho_s}\mbf{ \nabla}\delta p
    -\frac{1}{\rho_s}\delta\rho\,\zhat
    +E\nabla^2{\bf v},
\ee
or in terms of the individual cylindrical components,
\bea
  E^{1/2} \frac{\partial u}{\partial t}-v&=&-\frac{\partial}{\partial r}
   \frac{\delta p}{\rho_s}
   +E\left(\nabla^2-\frac{1}{r^2}\right)u  \label{vrad}  \\
  E^{1/2}\frac{\partial v}{\partial t}+u
    &=&E\left(\nabla^2-\frac{1}{r^2}\right)v \label{vtheta} \\
  E^{1/2}\frac{\partial w}{\partial t}&=&-\frac{1}{\rho_s}
   \frac{\partial}{\partial z}\delta p
    -\frac{\delta\rho}{\rho_s}+E\nabla^2 w \label{vvert},
\eea
where $(u,v,w)$ are the velocities in the $(r,\theta,z)$
directions.
We need two more equations in order to complete
the formulation of the problem, an equation of state
and the continuity equation.


We describe the fluid
in terms of the pressure and the concentrations of its
constituent elements. Within the context of neutron stars these elements are
mainly electrons, protons and neutrons. The equation of state then relates the
density to these quantities,
$\rho_*=\rho_* (p_*, Y_i)$ where $Y_i$ is the concentration of
the $i$-th particle species\footnote {In the core of an equilibrium neutron
star the
$Y_i$  are the concentrations that minimize the free energy through nuclear and
 weak interaction reactions. In the perturbations considered here, the
fluctuation time scales are short compared to those for the weak interactions
to adjust the ratio of neutrons to proton. The values of the $Y_i$ can thus
be considered as fixed properties of the matter. If the
equilibrium values of $Y_i$ give a stable stratification,
buoyant forces will cause perturbations to oscillate with the
Brunt--V\"ais\"al\"a frequency.}.

The nature of the restoring force and the corresponding
Brunt--V\"ais\"al\"a frequency is most readily calculated in  the
Lagrangian, as opposed to the 
Eulerian, formulation of the perturbations. 
We use $\delta q_*$ for an Eulerian perturbation of a 
quantity $q_*$, the difference between the actual and 
non-perturbed values of that quantity at a given point in space
and time. A
Lagrangian perturbation $\Delta q_*$ describes the change from the 
non-perturbed value an element of fluid experiences as it travels
{}from one point to another. The two perturbations are related by a
displacement vector field, $\mbf{\xi}_*$,
\be
  \Delta q_*=\delta q_* +
  \mbf{\xi}_*\cdot\mbf{\nabla}_* q_{0*},  
    \label{perts}
\ee
where $q_{0*}({\bf r})$ is the non-perturbed quantity.
The displacement vector field $\mbf{\xi}_*$ 
is related to the velocity by,
\be
  {\bf v}_*=\ddts\mbf{\xi}_*.
\ee
In non-dimensional notation
$\mbf{\xi}_*=(L_*\Delta\Omega_*/2\Omega_*)\,\mbf{\xi}$ and
\be
  {\bf v}=E^{1/2}\ddt\mbf{\xi}.
\ee

To relate the density and pressure perturbations,
consider a fluid displacement in which
some quantity $Y$ is held constant i.e.\ $\Delta Y=0$. The Lagrangian
perturbations $\Delta\rho_*$ and $\Delta p_*$ are then related by
\be
  \Delta\rho_*=\left(\frac{\partial\rho_*}{\partial p_*}\right
  )_Y\!\!\Delta p_*\equiv\frac{1}{c_{Y*}^2}\Delta p_*.
\ee
If the fluid is displaced adiabatically so that the entropy and
composition are fixed, then $c_{Y*}$ is the usual sound speed.
We characterize the 
equilibrium relationship between the density and the pressure
by
\be
  \frac{\partial\rho_{s*}/\partial z_*}{\partial p_{s*}/\partial z_*}
  =\left (\frac{\partial\rho_*}{\partial p_*}\right )_{\rm eq}
  \equiv\frac{1}{c_{\rm eq*}^2}. \label{ceq1}
\ee
%
With (\ref{perts})--(\ref{ceq1})
we can relate $\delta\rho_*$ and $\delta p_*$:
\bea
  \delta\rho_* &=& \Delta\rho_*-\mbf{\xi}_*\cdot
	          \mbf{\nabla}\rho_{s*} \\
   &=&  \frac{1}{c_{Y*}^2}\delta p_* + \left (\frac{1}{c_{{\rm eq*}}^2}-
    \frac{1}{c_{Y*}^2}\right )\rho_{s*} g_* \xi_{z*}, \label{drho1}
\eea
with $\xi_{z*}$ the $z$-component of $\mbf{\xi}_*$.
Once again, non-dimensionalizing we obtain,
\be
  \delta\rho=\left (\frac{g_*L_*}{c_{Y*}^2}\right)\delta p 
  + \left (\frac{N_*^2}{4\Omega_*^2}\right)\rho_s\xi_z
  \equiv \ky\delta p+N^2\rho_s\xi_z,  \label{drho}
\ee
where  the Brunt--V\"ais\"al\"a frequency\ $N_*$ is
\be
  N_*^2\equiv g_*^2\left(\frac{1}{c_{\rm eq*}^2}-\frac{1}{c_{Y*}^2}\right ),
\ee
and the two dimensionless parameters $\ky\equiv
g_*L_*/c_{Y*}^2$ and $N=N_*/2\Omega_*$ are the
``constant-$Y$ compressibility'' and the normalized
Brunt--V\"ais\"al\"a frequency, respectively. In previous studies
$\ky$ was assumed to be negligible, but in self gravitating
astronomical bodies $\ky$ can be of order unity or much larger.
Returning to our example of the neutron star, for instance, we
can estimate the size of $\ky$. Using the values
$g_*\approx10^{14}{\rm cm/sec^2}$, $L_*\approx10^6{\rm cm}$ and
$c_{Y*}\approx 10^9{\rm cm/sec}$ (Epstein 1988), we obtain
$\ky\approx10^2$.  $N$ characterizes the influence of
density stratification on Ekman pumping. 

At this point, it is worth noting that (\ref{drho1}) is a
generalization of more familiar formulas for low frequency density
perturbations in a compressible stratified fluid. In particular,
in studies of the terrestrial atmosphere one usually
assumes that the fluid motions are slow enough to allow
the pressure of displaced fluid to adjust to its surroundings
so that $\delta p=0$ and the first term in the left hand side of
(\ref{drho1}) vanishes.
This is equivalent to what we find in \S 6.4 of Pedlosky
(1979). Furthermore, if we consider the incompressible limit,
$c_{Y*}^2\rightarrow \infty$, then
\be
  N_*^2\rightarrow-\frac{g_*}{\rho_*}\frac{\partial\rho_*}{\partial z_*},
\ee
which is the familiar incompressible formula for the
Brunt--V\"ais\"al\"a frequency.

The final equation is the continuity equation for the
perturbations, 
\bea
  \frac{\Delta\rho_*}{\rho_{s*}}&=&-\mbf{\nabla}_*\cdot\mbf{\xi}_* \\
  &=&-\nabla_{r*}\cdot\xi_{r*}-\frac{\partial\xi_{z*}}{\partial z_*}.
\eea
With  (\ref{perts}), the continuity equation becomes,
\be
  \delta\rho_*+\xi_{z*}\frac{\partial\rho_{s*}}{\partial z_*}+\rho_{s*}
    \nabla_{r*}\cdot\xi_{r*}+\rho_{s*}\frac{\partial\xi_{z*}}{\partial z_*} =0.
  \label{conta}
\ee
Using  (\ref{ceq1}) and (\ref{drho1}), and taking
the time derivative of (\ref{conta}), we get
\be
  \frac{1}{\rho_{s*} c_{Y*}^2}\frac{\partial}{\partial t_*}\delta p_*
   - \frac{g_*}{c_{Y*}^2}w_*+\nabla_{r*}\cdot u_*
   +\frac{\partial w_*}{\partial z_*}  =0.
\ee
In non-dimensionalized units this is
\be
  E^{1/2} \Omega^2\frac{\partial}
   {\partial t}\frac{\delta p}{\rho_s}-\ky w
   + \frac{\partial w}{\partial z} + \frac{1}{r}\frac{\partial}
   {\partial r}(ru) =0,  \label{cont}
\ee
where we have introduced the dimensionless angular velocity,
\be
  \Omega\equiv\frac{2L_*\Omega_*}{c_{Y*}}.
\ee

We can now rearrange the complete set of perturbation
equations in a more convenient form.
We use \ (\ref{drho}) to eliminate $\delta\rho$
in \ (\ref{vvert}) and take the time
derivative to obtain,
\be
  E\frac{\partial^2w}{\partial t^2}=-N^2w-E^{1/2}\ddz\ddt
   \frac{\delta p}{\rho_s}
   -E^{1/2}(\ky-\ke)\ddt
   \frac{\delta p}{\rho_s}
   +E^{3/2}\nabla^2 \ddt w. \label{vvert2}
\ee
where the ``equilibrium compressibility'' is
\be
  \ke\equiv\frac{g_*L_*}{c_{\rm eq*}^2}.
\ee
Since the constant-$Y$ and equilibrium compressibilities are
comparable, we write
\be
  \eps\equiv\ke-\ky=\frac{N_*^2L_*}{g_*}=FN^2\frac{L_*}{r_{c*}}\ll 1.
\ee
The variable $\delta p$ only
occurs in the combination $\delta p/\rho_s$, so we define
$\delta P\equiv\delta p/\rho_s$. The final equations are now
\be
  E^{1/2} \ddt u-v=-\ddr\delta P
   +E\left(\nabla^2-\frac{1}{r^2}\right)u  \label{eq1} \ee\be
  E^{1/2}\ddt v+u
   =E\left(\nabla^2-\frac{1}{r^2}\right)v \label{eq2} \ee\be
  E\frac{\partial^2w}{\partial t^2}=-N^2w-E^{1/2}\ddz\ddt\delta P
   +E^{1/2}\eps\ddt\delta P
   +E^{3/2}\nabla^2 \ddt w. \label{eq3} \ee\be
  E^{1/2} \Omega^2\ddt\delta P
   -\ky w
   + \ddz w + \frac{1}{r}\ddr(ru) =0.  \label{eq4}
\ee

Note the harmonic restoring force provided by the
Br\"unt--V\"ais\"al\"a term in (\ref{eq3}).
The above four equations, (\ref{eq1})--(\ref{eq4}), describe the
evolution of the four unknowns, ${\bf v}$ and
$\delta P$. We  have  dimensionless parameters,
$E$, $N^2$, $\eps$, $\ky$ and $\Omega^2$. In
order to reduce the parameter space we consider only slow
rotation, $\Omega^2\ll 1$, and $\eps\ll 1$. Both
$\Omega^2$ and $\eps$ are easily included in the general solution,
but since they only appear as a product with $E^{1/2}$ their
effects are small and we do
not consider these terms in what follows.
Table 1 contains  definitions of the
dimensionless parameters and 
representative values for a neutron star.

The presence of the compressibility $\ky$
distinguishes this set of equations from earlier studies (Walin
1969; Sakurai 1969; Clark {\it et al}.\ 1971). 
Previous studies chose to emphasize the effects of
temperature on the density of the fluid. Specifically, the
density was considered a function of the temperature and the
stratification was a result of a temperature gradient which was
imposed by the boundary conditions. The dynamical significance
of the stratification and Brunt--V\"ais\"al\"a frequency arose
through the effects of temperature diffusion and the heat
equation. This approach is not appropriate to
the astrophysical cases with
which we are primarily concerned. 
In neutron stars, for example, thermal effects have a negligible
result on the fluid dynamics, whereas compressibility is quite
significant. We, therefore, focus on the dependence on the
equation of state.

\begin{center}
{\it 2.2 Boundary values and initial conditions}
\end{center}

To obtain a unique solution to 
(\ref{eq1})--(\ref{eq4}), we need to specify both the boundary
and initial conditions to our problem. There are, in essence, two
approaches to take at this point. The most complete method 
is to state that initially the fluid rotates uniformly with the
cylinder, and solve for the  behaviour of the fluid
after the angular velocity of the cylinder changes 
with the Laplace transformation technique
(Greenspan \& Howard 1963).
A simpler, and more physically elucidating approach, although less
rigorous,  used by other researchers in the
field (Walin 1969; Sakurai 1969; Barcilon \& Pedlosky 1967)
entails  recognizing that different
physical processes take place on widely different time
scales in different
regions of the fluid. We will follow this latter approach.

If the Ekman number $E$, or dimensionless viscosity, is
sufficiently small, the behaviour of the fluid following an
abrupt change in rotation rate of the container can be viewed as
three distinct physical processes which occur on time scales
$\Omega_*^{-1}$, $E^{-1/2}\Omega_*^{-1}$ and
$E^{-1}\Omega_*^{-1}$. The most rapid process is the formation
of a viscous boundary layer.
Following the impulsive change of rotation of the cylinder,
a viscous Rayleigh shear layer forms on the upper and lower
surfaces in a time scale on
the order of a rotation time ($t_{b*}\approx\Omega_*^{-1}$). Within this
region the gradient in the
azimuthal velocity results in an imbalance between the
centrifugal and pressure gradient forces causing fluid to
flow radially. This radial flow in the boundary layer
establishes a secondary flow where fluid in 
the interior is pulled into the boundary layer to replace the
flow in the Ekman layer, creating an opposing radial
flow in the interior fluid that satisfies continuity
requirements. This Ekman pumping spins the interior of the fluid up
in a time scale of order $E^{-1/2}\Omega_*^{-1}$. 
With our choice of dimensionless variables this corresponds
to a dimensionless
time, $t_E\approx1$. Finally, residual
oscillations decay in the viscous diffusion time $t_{v*}\approx
E^{-1}\Omega_*^{-1}$.

Since the principal goal of this investigation is to understand the
effects of the stratification and compressibility on the
Ekman pumping in the interior of the fluid, we 
expand  (\ref{eq1})--(\ref{eq4}) in powers of
$E^{1/2}$ and isolate the equations relating to Ekman pumping.
The initial velocity distribution
for the Ekman pumping equation is equivalent to the final velocity
distribution of the boundary layer which 
forms during the first phase. 
Following Walin (1969), we formulate the boundary
condition in terms of the continuity of the velocity
perpendicular to the Ekman boundary just outside the boundary
layer,
\bea
  w(z=\pm 1)&=&\mp\frac{E^{1/2}}{\sqrt 2}(\mbf{\nabla}\times{\bf v})_z \\
   &=&\mp\frac{E^{1/2}}{\sqrt 2}\frac{1}{r}\frac{\partial}{\partial r}
   (rv).  \label{bc}
\eea

It is critical to the  dynamics of Ekman pumping that the vertical velocity at
the boundary layer is
$O(E^{1/2})$.  This standard result  (see, e.g.  Pedlosky 1979) can be
understood by scaling arguments. The imbalance between the
centrifugal forces and pressure gradient forces  in the boundary layer drives
the Ekman pumping process. The thickness $\lambda$ of the boundary layer is
$O(E^{1/2})$ since the viscous terms in the dimensionless Navier--Stokes
equation is $E\nabla^2  \approx E/\lambda^2 = O(1)$. The mass flux {\it within} the
boundary layer is $\dot M_\lambda \propto \lambda = O(E^{1/2})$. The net mass
flux $\dot M_z \propto w$ {\it perpendicular} to the boundary layer is of the
same order as
$\dot M_\lambda$ giving $w = O(E^{1/2})$.
The sidewalls also have an $O(E^{1/2})$ boundary layer. As shown
in Pedlosky (1967) the vertical velocity through this layer is
inhibited by buoyancy forces and is
$O(E^{1/2})$ leading to a net mass flux of $O(E)$. The
boundary condition of the radial velocity is then $u=O(E)$ at
$r_c$. Since we keep terms only up to $O(E^{1/2})$, as described
in the following section, this gives
$u(r_c)=0$.

\section{Solutions}

To solve  (\ref{eq1})--(\ref{eq4})
perturbatively we expand each fluid variable $q$
as a series $q=q_0+E^{1/2}q_1+Eq_2+\cdots$.
Collecting terms of a
given power of $E^{1/2}$, we obtain a set of
equations governing each order in the expansion. 

We find that the $O(1)$ 
equations are,
\be
  v_0=\ddr\delta P_0 \label{b1} \ee\be
  u_0=0 \ee\be
  w_0=0 \ee\be
  -\ky w_0+\ddz w_0 +\frac{1}{r}\ddr(ru_0)=0,
\ee
and the $O(E^{1/2})$ equations are,
\be
  v_1=\ddr\delta P_1 \ee\be
  \ddt v_0=-u_1 \label{b2} \ee\be
  N^2w_1=-\ddz\ddt\delta  P_0  
                             \label{b3}  \ee\be
  -\ky w_1+\ddz w_1+\frac{1}{r}\ddr(ru_1)=0.  \label{b4}
\ee
We define $\phi\equiv-\partial\delta P_0/\partial t$, so that 
 (\ref{b1}) and (\ref{b2}) become
\be
  u_1=\ddr\phi  \label{b4.1}
\ee
and (\ref{b3}) and (\ref{b4}) are now, respectively,
\be
  N^2w_1=\ddz\phi  
                    \ee\be
  \frac{-\ky}{N^2}\left(\ddz\phi 
  \right)+\ddz\frac{1}{N^2}
    \left(\ddz\phi 
    \right)+\frac{1}{r}\ddr(r\ddr\phi)=0. \label{b5}
\ee
Assuming $N^2$ varies slowly 
over $z$, we treat it as a constant
and simplify (\ref{b5}) to
\be
   \frac{\partial^2}{\partial z^2}\phi-\ky 
   \ddz\phi+
   N^2\frac{1}{r}\ddr
    \left(r\ddr\phi\right)=0.  \label{b6}
\ee
After taking the time derivative, the boundary condition,
(\ref{bc}), is 
\be
  \ddt\ddz\phi 
  =\pm\frac{N^2}{\sqrt2}\frac{1}{r}\ddr
    \left(r\ddr\phi\right),{\rm\ \ at\ }z=\pm 1.
\ee
By setting $\phi=Z(z)\,R(r)\,T(t)$,  (\ref{b6}) becomes,
\be
  \frac{1}{Z}\frac{d^2}{dz^2}Z-\frac{\ky 
  }{Z}\frac{d}{dz}Z
   +\frac{N^2}{R}\frac{1}{r}
   \frac{d}{dr}\left(r\frac{d}{dr}R\right)=0.
\ee
The solutions to the spatial functions are
\bea
&&  Z=A{\rm e}^{\bp z}+B{\rm e}^{\bm z} \\
&&  R=J_0(kr), \label{bessel}
\eea
where 
\be
 \beta_\pm=\frac{1}{2}\left( 
   \ky\pm(\ky 
    ^2+4k^2N^2)^{\frac{1}{2}}\right)\label{b7}
\ee
The symmetry of the boundary condition,
$w_1(z=0)=0$, relates $A$ and $B$:
\be
  A=-\frac{\bm 
    }{\bp 
    }B.
\ee
The constant $B$ is arbitrary, and we
choose it so that $Z(1)=1$. This leads to
\be
  B=\left({\rm e}^\bm-\frac{\bm 
    }{\bp 
    }{\rm e}^\bp\right)^{-1}.
\ee
The possible values of $k$ are determined by the boundary
condition at the sidewall at $r=r_c$, i.e.\ $u_1(r_c)=0$. 
{}From (\ref{b4.1}) and (\ref{bessel}) we see that this condition
corresponds to $J_1(k_mr_c)=0$ for $m=0,1,2,\ldots$. The first
zeros of $J_1$ are $k_mr_c=0,\ 3.8317,\ 7.0156,\ldots$.
The solution $k_m=0$ has ${\bf v}=0$ everywhere and is of no
interest. 

We  utilize the boundary condition to determine the time dependence
of $\phi$. Putting our solution for $R$ and $Z$ into \
(\ref{bc}), we obtain the differential equation,
\be
  \frac{d}{dt}T=\frac{-k^2N^2}{\sqrt2}\frac{Z(1)}{\frac{dZ}{dz}(1)
    }T,
\ee
whose solution is,
\be
  T(t)={\rm e}^{-\omega t},
\ee
where,
\be
  \omega=\frac{k^2N^2}{\sqrt2}\left(A\bp {\rm e}^\bp+B\bm {\rm e}^\bm 
    \right)^{-1},
\ee
with $A$ and $B$ defined as above.

There are two interesting limiting cases. The first is that of
no stratification $N\rightarrow0$; the second is that of an
incompressible fluid $\ky\rightarrow0$. Let us
consider the first of these which gives,
\be
  \bp\approx \ky\left(1+\frac{k_m^2N^2}{\ky^2}\right) \ee\be
  \bm\approx -\frac{k_m^2N^2}{\ky} \ee\be
  B\approx 1+\frac{k_m^2N^2}{\ky}(1+{\rm e}^\ky/\ky)\ee\be
  A\approx \frac{k_m^2N^2}{\ky^2}
\ee
and,
\be
  \omega\approx\frac{\ky}{\sqrt2 ({\rm e}^\ky-1)}.\label{omega}
\ee
Equation (\ref{omega}) shows that for large compressibilities
the Ekman spin-up time scale 
$\omega^{-1}$ grows exponentially with $\ky$.
The second limiting case,
$\ky\rightarrow0$, gives
$\beta_\pm\approx\pm k_mN$ and 
\be
  \omega\approx\frac{k_mN\cosh{k_mN}}{\sqrt2\sinh{k_mN}}.
\ee
This matches the $\ky=0,\,N\ne0 $ solution which was obtained by
Walin (1969). 

We are now in a position to write the complete solution for the
quantities $\phi$ and $v_0$;
\be
  \phi=\sum_{m=1}^\infty C_mZ_m(z)J_0(k_mr){\rm e}^{-\omega_mt} .
\ee
The velocity, $v_0$, is found from the relationship,
\be
  \ddr\phi=-\ddt v_0,
\ee
which gives,
\be
  v_0(r,z,t)=-\sum_{m=1}^\infty \frac{k_m}{\omega_m}C_mZ_m(z)J_1(k_mr)
    {\rm e}^{-\omega_mt}+v_\infty(r,z) .
\ee
The last term represents the final velocity due to Ekman pumping.
If we take the frame of reference as that rotating with
the cylinder before  impulsive
spin-up, the 
final velocity at the boundary of the interior fluid is
\be
  v_\infty(r,z=\pm 1)=r.
\ee
We determine $C_m$ from the initial state of the fluid,
\be
  v_0(r,z=1,t=0)=0=-\sum_{m=1}^\infty\frac{k_mC_m}{\omega_m}J_1(k_mr)+r.
\ee
The coefficients, given by the standard equation for 
a Fourier--Bessel series, are
\be
  \frac{k_mC_m}{\omega_m}=\frac{2}{k_mJ_2(k_m)},
\ee
and the final velocity is 
\be
  v_\infty(r,z)=2\sum_{m=1}^\infty\frac{1}{k_mJ_2(k_m)}J_1(k_mr)
    Z_m(z),
\ee
where we have chosen $r_c=1$ ($r_{c*}=L_{*}$).

\section{Discussion}

The time dependence of the Ekman pumping process is exponential with
characteristic time $1/\omega$. We plot the value of $\omega$
as a function of $k_mN$ in figure\ 1 for different values of the
parameter $\ky$. 
Larger $N$, corresponding to greater density stratification
gives  larger $\omega$ and reduced characteristic time. That is,
a strongly stratified fluid spins up
much quicker than a non-stratified fluid.
On the other hand, an increased value of the compressibility
$\ky$ slows the pumping process for a 
given  $k_mN$. The spin-up time $\omega^{-1}$ decreases with increased
stratification because stratification  isolates much of the
fluid from the pumping process. 

Figures 2 and 3 show that
the rotation state at the end of the Ekman pumping stage is not
that of a solid body. 
The ordinate $Z(z)$ is proportional to
the final azimuthal velocity, with $Z=1$ being the largest
possible spin-up.
Larger values of $k_mN$ leave more of the internal fluid
unaffected by the Ekman pumping process. In contrast, in a
homogeneous fluid, $N=0$, Ekman pumping brings the entire fluid
to an 
angular velocity equal to that of the boundary.
The compressibility $\ky$ further
decreases the amount of pumped fluid, as we can see by comparing
figure\ 2,  for $\ky=0$, with figure\ 3,  for
$\ky=10$. 

Compressibility thus decreases the efficacy of Ekman pumping
both by lengthening the spin-up time and by decreasing the
amount of affected fluid.
To convey a clearer picture of how strong the effect of $\ky$ is,
we plot in figure 4 the final angular velocity of the fluid at its
central ($z=0$) layer as a function of $k_mN$ for different values
of $\ky$. Though there is little
change between $\ky=0$ and $\ky=1$, the internal final angular
velocity is strongly suppressed as $\ky$ increases to $10$.

We point out that for canonical values $N^2\approx6$ and
$\ky\approx10^2$ for a neutron star, the thickness of the layer
affected by Ekman pumping is much smaller than the
radius of the cylinder. Gradients in the gravitational
acceleration are
therefore small in these layers, justifying our original
assumption of constant $g_*$ in (2).

In figure\ 5 we plot the average spin-up of the fluid $\langle
Z\rangle$ as a function of the normalized Brunt--V\"ais\"al\"a
frequency $N$ for the two lowest order modes, $k_1$ and $k_2$.
We see that even modest
values of $N$ prevent most of the fluid from spinning up during
the Ekman pumping phase.
The state of the fluid after a 
time scale of $t_*\approx E^{-1/2}\Omega_*^{-1}$
is, thus, one of non-uniform rotation. The process
of viscous diffusion, which operates in a time
$t_{v*}\approx E^{-1}\Omega_*^{-1}$ eventually brings the fluid into 
solid-body rotation.

The case of spherical geometry was studied by Clark {\it et al}.\ 
(1971), where they found that the solution for a sphere is
qualitatively similar to that of the cylinder. That is, the
final state of non-uniform rotation also exists in the sphere,
but the geometry of the layer that gets Ekman pumped is
modified.

We find a particularly interesting application of  these
phenomena is the response of the interior of a rotating neutron
star to a glitch, a sudden small change in the rotational
velocity. Within the star there exists a significant
stratification due to the strong gravitational field and the
equilibrium concentrations of protons, neutrons and electrons.
Reisenegger \& Goldreich (1992) estimated a value of
$N_*\approx500{\rm s}^{-1}$ for a neutron star. For a canonical
value of $\Omega_*\approx100{\rm s}^{-1}$, we obtain
$N\approx2.5$.
This is large enough to have a significant effect
on the length of time the core of the star needs to come into
rotational equilibrium. We explore these issues in a forthcoming
paper. 

We would like to thank Angela Olinto for her valuable input and
advice. 
M.A. was supported in part by the DOE at Chicago, by NASA
grant NAG 5-2788 at Fermi National Laboratory, and through a
collaborative research grant from IGPP/LANL. This work was
carried out under the auspices of the U.S. Department of Energy.

\bigskip

\noindent
\large{\bf References}
\normalsize

\begin{description}

\item Barcilon, V. \& Pedlosky, J. 1967, Linear theory of
rotating stratified fluid motions. {\it J. Fluid Mech.} {\bf
29}, 1.

\item Bark, F. H., Meijer, P. S. \& Cohen, H. I. 1978, Spin up
of a rapidly rotating gas. {\it Physics of Fluids} {\bf 21},
531--539.

\item Benton, E. R. \& Clark, A. 1974, Spin Up. {\it Ann. Rev.
Fluid Mech.} {\bf 6}, 257--280.

\item Buzyna, G. \& Veronis, G. 1971, Spin-up of a stratified
fluid: theory and experiment. {\it J. Fluid Mech.} {\bf 50},
579--608. 

\item Clark, A., Clark, P. A., Thomas, J. H. \& Lee N.
1971, Spin-up of a strongly stratified fluid in a sphere. {\it
J. Fluid Mech.} {\bf 45}, 131--149.

\item Epstein, R. I. 1988, Acoustic properties of neutron stars.
{\it Astrophys. J.} {\bf 333}, 880--894.

\item Greenspan, H. P. \& Howard, L. N. 1963, On a
time-dependent motion of a rotating fluid. {\it J. Fluid Mech.}
{\bf 17}, 385--404.

\item Howard, L. N., Moore, D. W. \& Spiegel, E. A. 1967, Solar
spin-down problem. {\it Nature} {\bf 214}, 1297--1299.

\item Hyun, J. M. 1983, Axisymmetric flows in spin-up from rest
of a stratified fluid in a cylinder. {\it Geophys. Astrophys.
Fluid Dyn.} {\bf 23}, 127--141.

\item Hyun, J. M. \& Park, J. S. 1990, Early time behavior of
the Ekman layers in spin-up of a rapidly rotating gas. {\it J.
Phys. Soc. Japan} {\bf 59}, 3584--3594.

\item Hyun, J. M. \& Park, J. S. 1992, Spin-up from rest of a
compressible fluid in a rapidly rotating cylinder. {\it J. Fluid
Mech.} {\bf 237}, 413--434.

\item Lindblad, I. A. A., Bark, F. H. \& Zahrai, S. 1994,
Spin-up of a rapidly rotating heavy gas in a thermally insulated
annulus. {\it J. Fluid Mech.} {\bf 274}, 383--404.

\item Linden, L. N. \& van Heijst, G. J. F. 1984, Two-layer spin
up and frontogenesis. {\it J. Fluid Mech.} {\bf 143}, 69--94.

\item O'Donnell, J. \& Linden, P. F. 1992, Spin-up of a
two-layer fluid in a rotating cylinder. {\it Geophys. Astrophys.
Fluid Dyn.} {\bf 66}, 47--66.

\item Park, J. S. \& Hyun, J. M. 1994, Dynamical structure of
compressible fluid flows in an abruptly rotating cylinder. {\it
J. Phys. Soc. Japan} {\bf 63}, 528--535.

\item Pedlosky, J. 1967, The spin-up of a stratified fluid. {\it
J. Fluid Mech.} {\bf 28}, 463--479.

\item Pedlosky, J. 1979, {\it Geophysical Fluid Dynamics},
Springer.

\item Reisenegger, A. \& Goldreich, P. 1992, A new class of
g-modes in neutron stars. {\it Astophys. J.} {\bf 395},
240--249.

\item Rieutord, M. 1992, Ekman circulation and the
synchronization of binary stars. {\it Astron. Astrophys.} {\bf
259}, 581--584.

\item Sakurai, T. 1969, Spin down problem of a rotating
stratified fluid in thermally insulated circular cylinders. {\it
J. Fluid Mech.} {\bf 37}, 689--699.

\item Sakurai, T., Clark, A. \& Clark, P. A. 1971, Spin-down of
a Boussinesq fluid of small Prandtl number in a circular
cylinder. {\it J. Fluid Mech.} {\bf 49}, 753--773.

\item Sakurai, T. \& Matsuda, T. 1974, Gasdynamics of a
centrifugal machine. {\it J. Fluid Mech.} {\bf 62}, 727--736.

\item Spence, G. S. M., Foster, M. R. \& Davies, P. A. 1992, The
transient response of a contained rotating stratified fluid to
impulsively started surface forcing. {\it J. Fluid Mech.} {\bf
243}, 33--50.

\item Tassoul, J.-L. \& Tassoul, M. 1990, A time dependent model
for synchronization in close binaries. {\it Astrophys. J.} {\bf
359}, 155--163.

\item Tassoul, M. \& Tassoul, J.-L. 1992, On the efficiency of
Ekman pumping for synchronization in close binaries. {\it
Astrophys. J.} {\bf 395}, 604--611.

\item Walin, G. 1969, Some aspects of a time-dependent
motion of a stratified rotating fluid. {\it J. Fluid Mech.} {\bf
36}, 289--307.

\end{description}

\pagebreak

\noindent
\large{\bf Figure Captions}
\normalsize

\medskip

{\bf Figure 1}. The spin-up characteristic time, $\omega$, as a
function of $k_mN$ for varying values of $\ky$. 

\medskip

{\bf Figure 2}. The final azimuthal velocity as a function of
depth for an arbitrary value of the radius. A value of $Z=1$ is
complete spin-up, while $Z=0$ is no spin-up, with $\ky=0$.

\medskip

{\bf Figure 3}. As in figure\ 2, but with $\ky=10$.

\medskip

{\bf Figure 4}. The final velocity of the central layer of the
fluid ($z=0$), as a function of $k_mN$, for different values of
$\ky$.

\medskip

{\bf Figure 5}. The average final spin-up of the fluid as a
function of the stratification. The top two curves (solid and
dashed lines) were
calculated for an incompressible fluid, $\ky=0$, while for the
bottom two curves (dotted and dash--dotted lines) $\ky=10$. With
a highly compressible fluid 
($\ky=10$) even very small values of $N$ result in very little
spin-up from Ekman pumping.

\pagebreak

\begin{center}
\begin{tabular}{|c|c|c|} \hline
  parameter  &  formula  &  value \\ \hline
    $ E$  & $\nu_*/(2\Omega_*L_*^2)$ & $10^{-7}$ \\ \hline
     $ N^2 $&$ N_*^2/(4\Omega_*^2) $&$ 6$ \\ \hline
     $\ky$  &$ g_*L_*/c_{Y*}^2$ & $10^2$ \\ \hline
     $F$  &$ 4\Omega_*^2r_{c*}/g_*$ &$ 10^{-4}$ \\ \hline
     $\eps $& $N_*^2L_*/g_*$ & $10^{-4}$ \\ \hline
     $\Omega^2$ & $2\Omega_*^2L_*^2/c_{Y*}^2$ & $10^{-2} $ \\ \hline
\end{tabular}
\end{center}

\medskip

{\bf Table 1}. The dimensionless parameters. The values quoted
are order of magnitude estimates for a characteristic pulsar.

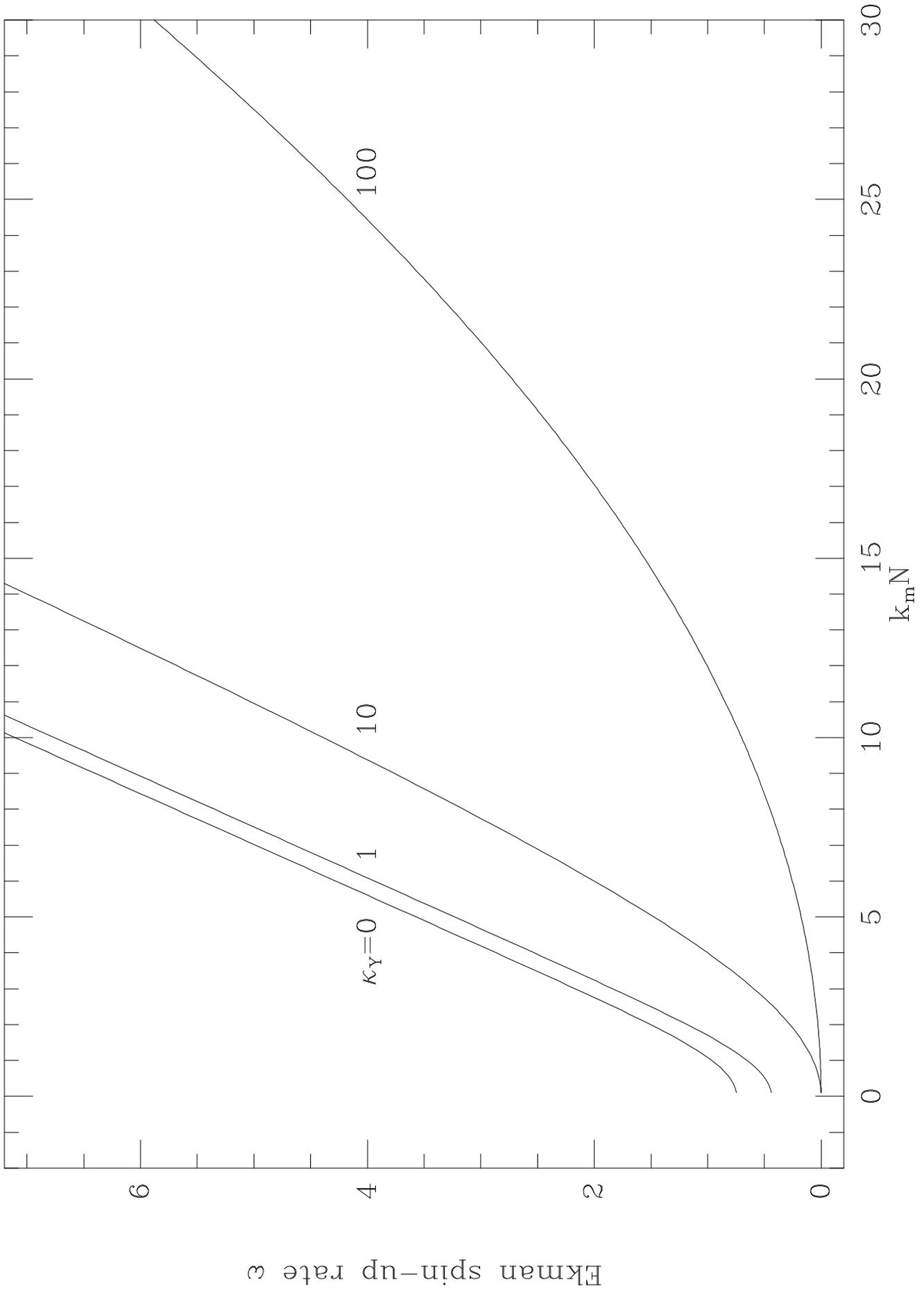
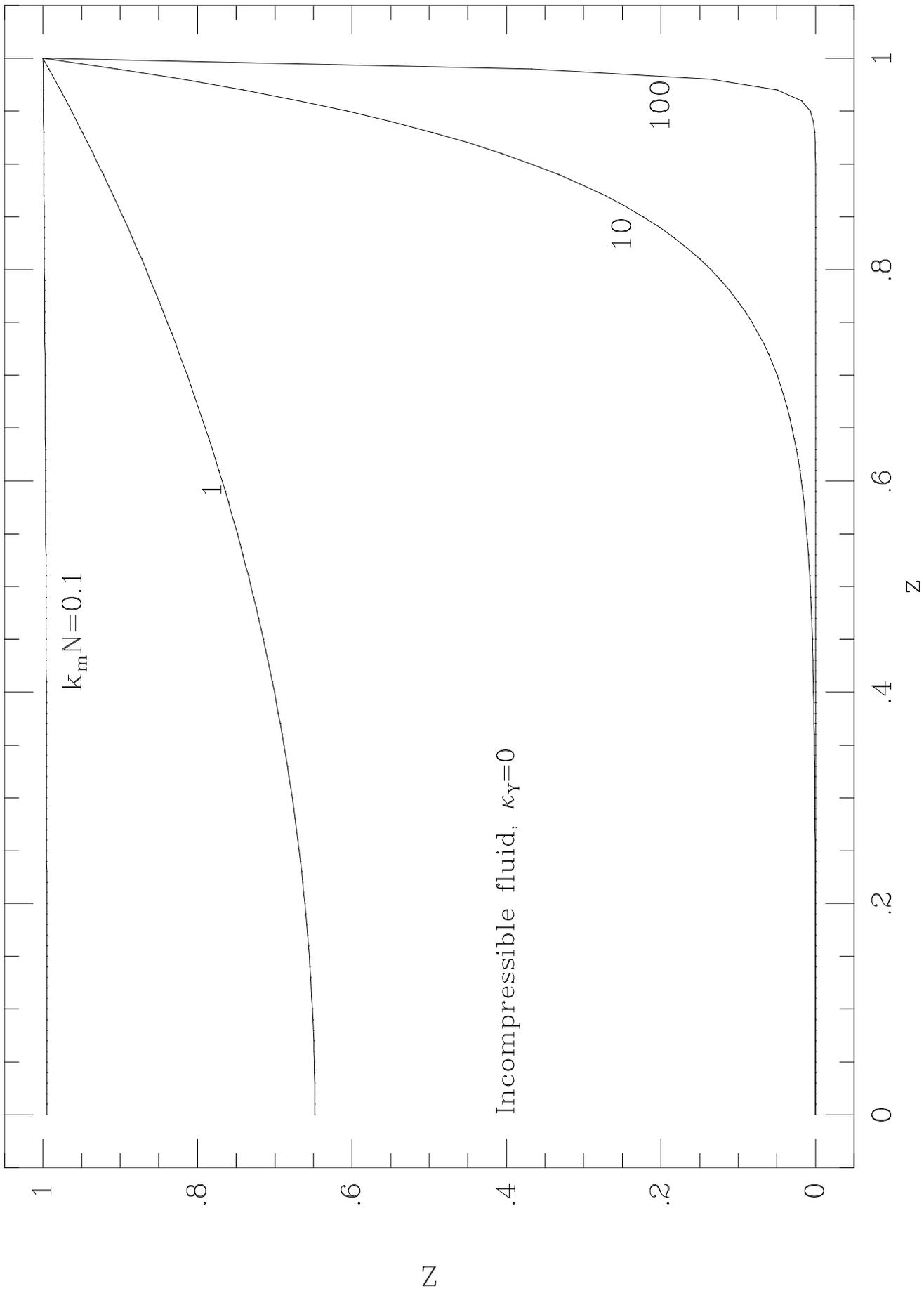
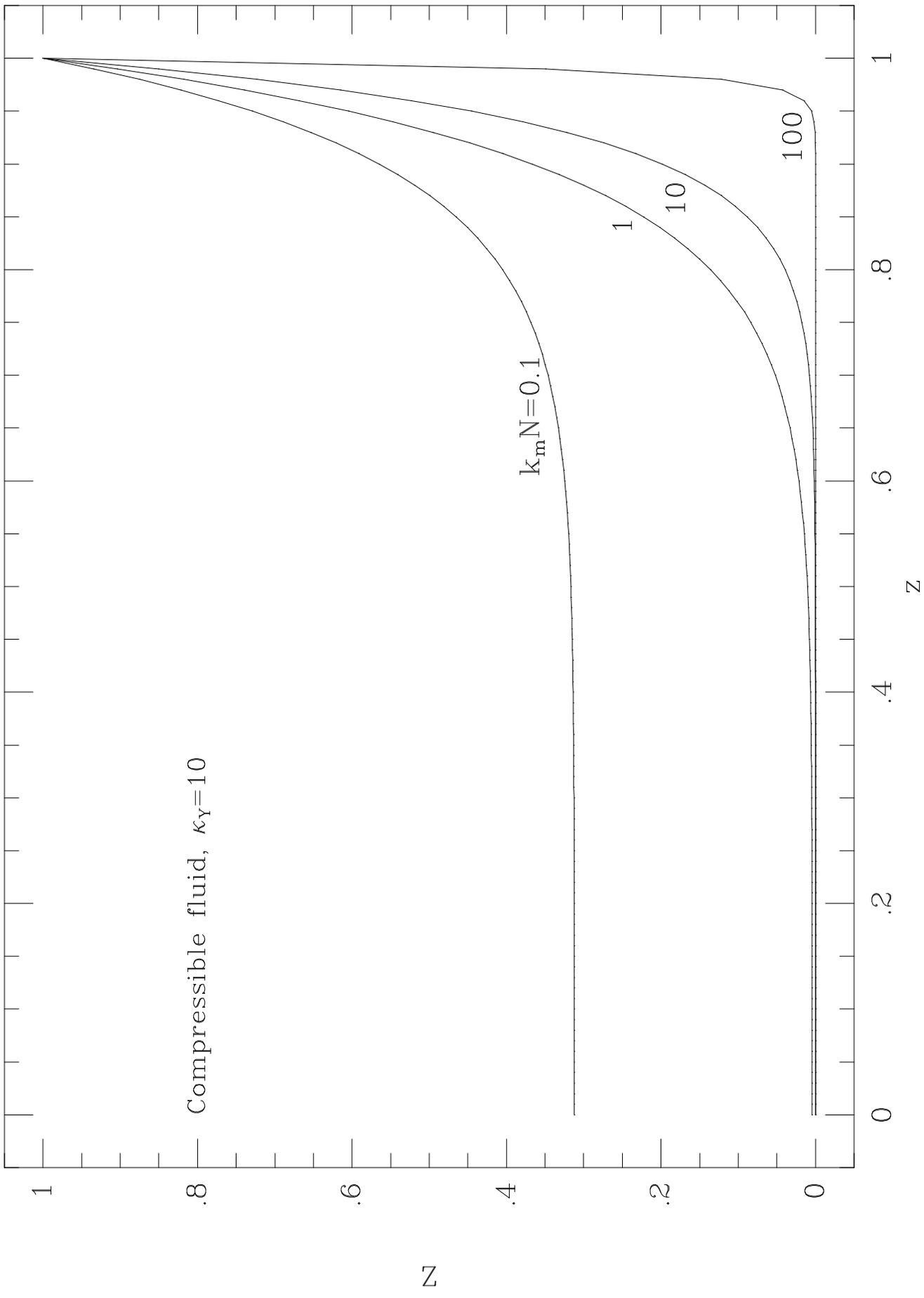
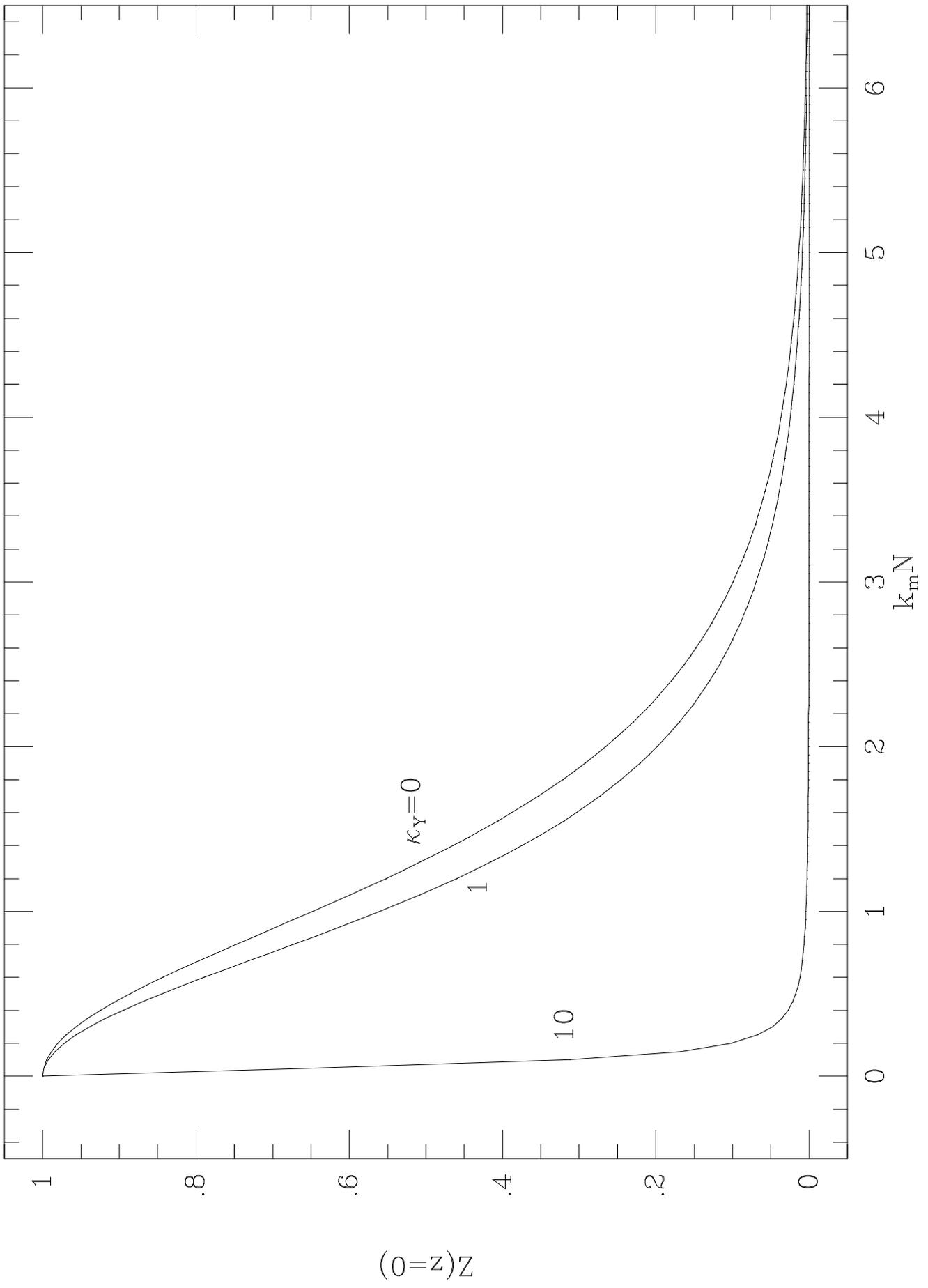
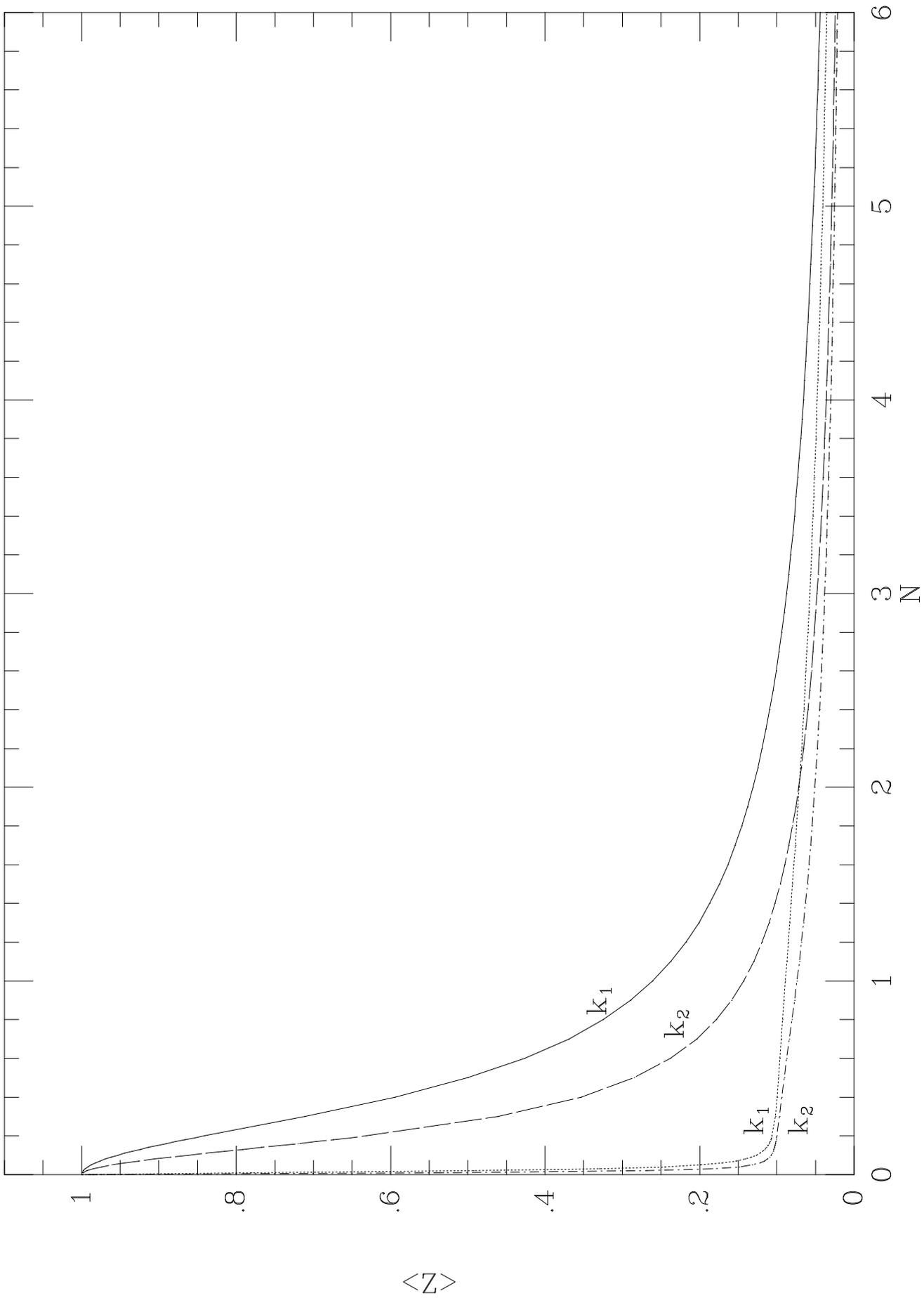
\end{document}